# Knowledge graph-based personalized multimodal recommendation fusion framework

**Author Yu Fang [1]**

**School of Chemistry and Chemical Engineering, Huazhong University of Science and Technology, Wuhan 430074, Hubei, China**

**E-mail: Yufang@hust.edu.cn**

**Abstract：**

In the contemporary age characterized by information abundance, rapid advancements in artificial intelligence have rendered recommendation systems indispensable. Conventional recommendation methodologies based on collaborative filtering or individual attributes encounter deficiencies in capturing nuanced user interests. Knowledge graphs and multimodal data integration offer enhanced representations of users and items with greater richness and precision. This paper reviews existing multimodal knowledge graph recommendation frameworks, identifying shortcomings in modal interaction and higher-order dependency modeling. We propose the Cross-Graph Cross-Modal Mutual Information-Driven Unified Knowledge Graph Learning and Recommendation Framework (CrossGMMI-DUKGLR), which employs pre-trained visual-text alignment models for feature extraction, achieves fine-grained modality fusion through multi-head cross-attention, and propagates higher-order adjacency information via graph attention networks.

**Key word：** Multimodal recommendation, Knowledge graph, Cross-modal attention, Graph attention network, Mutual information

## 1. Introduction

In this era of information explosion, the proliferation of digital platforms and services has created an unprecedented volume of data across diverse domains. With billions of users actively engaging with e-commerce, social media, streaming services, and digital content platforms, users confronting vast amounts of products, news, or audiovisual content urgently require efficient and precise personalized recommendations [1-3]. The challenge of information overload has become particularly acute as the variety and velocity of content generation continue to accelerate exponentially. Recent studies indicate that users typically abandon their search after examining only a fraction of available options, highlighting the critical need for intelligent filtering and recommendation mechanisms.

Traditional collaborative filtering methods rely on user-item interactions but often overlook multiple explicit attributes of users or items. These methods, while foundational to recommendation systems, suffer from several inherent limitations including data sparsity, cold start problems, and the inability to leverage rich contextual information. Furthermore, they fail to capture the semantic relationships between items and cannot provide explanations for their recommendations, limiting user trust and system transparency. Meanwhile, knowledge graphs (KG) [4, 5] can structurally organize items with entities, attributes, and relationships, thereby providing semantic-level auxiliary information for recommendations. Knowledge graphs represent a paradigm shift in how we model and understand complex relationships in recommendation scenarios, offering a principled approach to encode domain knowledge, capture multi-hop reasoning paths, and enable interpretable recommendations. Moreover, multimodal data such as text and images are now widely available across e-commerce, social, and media platforms, presenting new opportunities for understanding user preferences and item characteristics. The richness of multimodal data provides complementary perspectives on items visual features capture aesthetic preferences, textual descriptions convey functional attributes, and user reviews reflect experiential feedback enabling a more holistic understanding of user-item compatibility.

The integration of multimodal information with knowledge graphs presents unique challenges [6,7]. First, different modalities contain complementary yet heterogeneous information requiring sophisticated fusion strategies. The heterogeneity manifests not only in data formats and feature spaces but also in the temporal dynamics and noise characteristics of each modality. For instance, visual features may be more stable over time while textual reviews exhibit temporal drift in sentiment and topics. Second, knowledge graphs exhibit complex topological structures with multi-hop relationships that traditional methods fail to exploit. The propagation of information through knowledge graph structures involves navigating sparse connections, handling missing links, and reasoning over implicit relationships that may span multiple intermediate entities. This complexity is compounded when different modalities provide conflicting or incomplete information about the same entities. Third, the computational complexity of processing large-scale multimodal knowledge graphs poses scalability challenges for real-world applications. Modern recommendation platforms must handle millions of users and items, with knowledge graphs containing billions of triples and multimodal features requiring significant storage and processing resources. The challenge is further exacerbated by the need for real-time inference to support interactive user experiences.

Despite recent advances in multimodal learning and knowledge graph-based recommendations, existing approaches face several limitations. First, most methods treat different modalities independently or use simple concatenation strategies, failing to capture cross-modal interactions. Second, the integration of knowledge graphs with multimodal features often relies on shallow fusion mechanisms that cannot fully exploit the complementary nature of different data sources. Third, personalization in multimodal settings typically focuses on user-level preferences without considering the dynamic nature of modal importance across different contexts and items.

To address these challenges, we propose CrossGMMI-DUKGLR, a comprehensive framework that synergistically combines multimodal feature learning, knowledge graph reasoning, and personalized fusion mechanisms. Our approach introduces several technical innovations: (1) a cross-modal attention mechanism that captures fine-grained interactions between modalities, (2) a deep knowledge graph embedding module that learns hierarchical representations of entities and relations, (3) a personalized fusion strategy that adaptively weights different modalities based on user profiles and item characteristics, and (4) an interpretable recommendation generation process that provides reasoning paths through the knowledge graph. These components work together to create a unified framework that significantly advances the state-of-the-art in personalized multimodal

## 1.1. Background

Multimodal recommendation systems leverage multiple information sources user historical behavior, textual descriptions, product images, and videos to achieve comprehensive representation of user and item characteristics. Recent advances in deep learning have enabled sophisticated feature extraction from raw multimodal data. However, effective fusion of these heterogeneous features remains challenging. The primary challenges in multimodal recommendation include: feature heterogeneity, semantic gap and computational complexity.

Knowledge graphs encode entities and relationships into graph structures, providing semantic support for complex relationships in recommendations. Graph neural networks (GNNs) have revolutionized KG-based recommendations by enabling end-to-end learning on graph structures. However, purely structural approaches struggle to process visual and textual signals directly.

## 1.2. Problem Description

Although existing methods have made some progress, they still exhibit numerous shortcomings when confronted with more demanding tasks. First, most approaches merely concatenate or weight visual/text features, failing to capture their correlations at fine-grained levels. Second, applying one or two layers of GNNs on the knowledge graph alone struggles to fully leverage multi-hop entity relationships, hindering the modeling of user interest chains. Furthermore, the absence of a unified and flexible multimodal fusion mechanism makes it difficult to dynamically allocate contribution weights among visual, textual, and structural information.

## 2. Related Work

### 2.1 Knowledge Graph-based Recommendation

Knowledge graphs have revolutionized recommendation systems by providing structured representations of domain knowledge. KGCN [8] applies graph convolutional networks on knowledge graphs to aggregate neighborhood information. RippleNet [9] propagates user preferences through the knowledge graph like ripples. KGAT [10] uses attention mechanisms to distinguish the importance of different neighbors in the knowledge graph.

More recent approaches have focused on learning better knowledge graph embeddings for recommendation. CKAN [11] combines collaborative filtering with knowledge graph embeddings through a unified neural architecture. KGIN [12] models user intents as combinations of knowledge graph relations. However, these methods primarily focus on single-modal scenarios and do not fully exploit the potential of knowledge graphs in multimodal settings.

### 2.2 Multimodal Fusion in Recommendations

Multimodal recommendation has attracted increasing attention due to its ability to leverage diverse data sources. Yang et al. proposed MMGCN [13], which uses graph convolutional networks to model user-item interactions across different modalities. Li et al. introduced MVAE [14], a variational autoencoder framework for multimodal recommendation that learns joint latent representations. However, these methods typically process modalities in isolation before fusion, missing important cross-modal correlations.

Recent works have explored more sophisticated fusion strategies. Liu et al. developed a hierarchical attention network that progressively fuses modalities at different semantic levels [15]. Chen et al. proposed a contrastive learning framework [16] that aligns representations across modalities while preserving modality-specific information. Despite these advances, most existing methods lack explicit mechanisms for modeling the relationships between different modalities and struggle with interpretability.

### 2.3 Entity Alignment and Cross-Graph Learning

Entity alignment aims to identify equivalent entities across different knowledge graphs, enabling knowledge sharing and integration. Traditional methods rely on structural similarity or attribute matching. Recent approaches like MIKG maximize mutual information across knowledge graphs for robust alignment but neglect multimodal information. The integration of entity alignment with recommendation systems remains largely unexplored, presenting opportunities for cross-graph knowledge transfer to enhance recommendation quality.

### 2.4 Personalized Fusion Techniques.

Combining the methods A and B, the relevant literature "Multimodal Fusion Framework Based on Knowledge Graph for Personalized Recommendation (Multi-KG4Rec) [17]" and "Maximizing Mutual Information Across Knowledge Graphs for Robust Entity Alignment (MIKG) [18]" were identified. These were compared and summarized based on their methodological characteristics, advantages and disadvantages, as well as existing shortcomings.

CrossGMMI-DUKGLR leverages PVAM and cross-attention to fuse text/image information within KGs, enhancing recommendation accuracy. However, it relies on large models, incurs high deployment costs, and is limited to single-image scenarios. MIKG employs BERT-GNN encoding and maximizes cross-KG mutual information via InfoNCE for entity alignment, but suffers from computational intensity, neglects low-frequency attributes, and degrades under heterogeneous structures. Both approaches neglect cross-graph multimodal fusion, efficient negative sampling, and noise robustness, highlighting the urgent need for a unified cross-graph, cross-modal self-supervised framework.

## 3. Review of Method A

Paper：《Multimodal fusion framework based on Knowledge Graph for personalized Recommendation》

Overall Assessment: Weak Accept

Method Focus: Multi-KG4Rec

Technical Depth: High

Implementation Quality: Good

Presentation: Clear and Comprehensive

### 3.1. Summary of Method A

The Multi-KG4Rec model, based on the "Multimodal Fusion Framework Based on Knowledge Graph for Personalized Recommendation," employs a core approach that primarily involves decomposing the original KG into subgraphs according to modalities such as text and vision. Pre-trained models like "Contrastive Language–Image Pre-training" (CLIP) are utilized to extract initial features from text and images, respectively. Bidirectional Cross-Modal Multi-Head Attention achieves fine-grained modal fusion; It employs Graph Attention Networks for higher-order neighbor information propagation within the KG. Item recommendations are then generated through ranking loss.

The actual implementation process can be summarized as shown in Fig. 1:

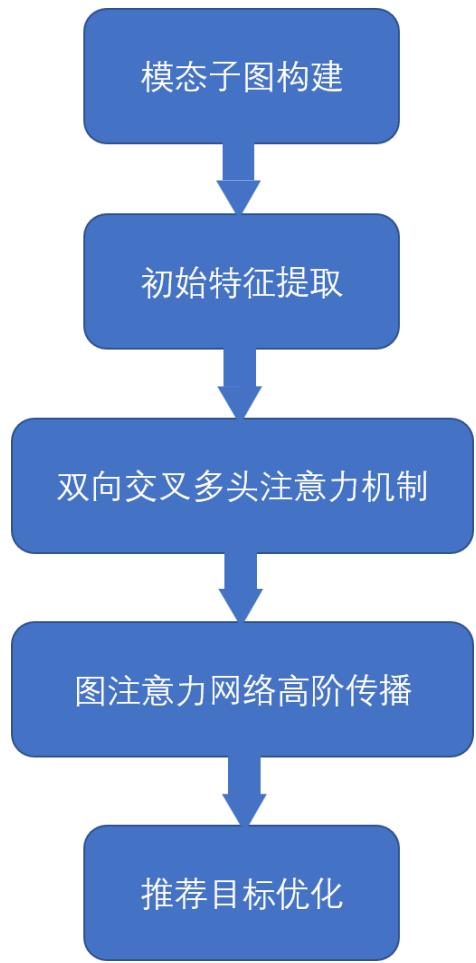

Fig.1. Implementation Process of Method A

## 3.2. Strong Points

**Strengths of BIMF-GAN Implementation (S1-S3)**

S1 Performs unified modeling across multiple data sources (graph, text, visual) to achieve fusion;

S2 Proposes a tripartite AutoEncoder for structural/visual/textual features, ensuring expressive capability in independent spaces;

S3 Achieves significant improvements on datasets including MovieLens and LastFM.

## 3.3. Weak Points

**Weaknesses of BIMF-GAN Implementation (W1-W3)**

W1 Linear fusion of three features occurs only after concatenation, lacking intermodal interaction;

W2 The autoencoder reconstruction task is decoupled from the recommendation objective, making it difficult to optimize representation contributions to downstream tasks;

W3 Limited support for propagating high-order graph neighbors, with a maximum of two GNN layers, failing to capture deep-level relationships.

## 3.4. Detailed Analysis

(1) CKE contributes insufficiently to downstream recommendation performance for autoencoder reconstruction tasks; introducing end-to-end multi-task training is recommended to unify objectives;

(2) Concatenation-based fusion limits modal interaction capabilities; attention mechanisms can be incorporated before and after concatenation to assign different weights to distinct modalities;

(3) Experiments only employed small-scale GNNs with 20 iterations, lacking sensitivity analysis on hyperparameters such as neighbor layer depth and neighbor sampling strategies.

## 4. Review of Method B

Paper：《Maximizing mutual information across Knowledge Graphs for robust Entity Alignment》

Overall Rating: Weak Accept

Method Focus: MIKG

Technical Depth: High

Implementation Quality: Good

Presentation: Adequate

## Summary of Method B

Based on the MIKG model from "Maximizing Mutual Information Across Knowledge Graphs for Robust Entity Alignment," this approach primarily targets cross-graph entity alignment. Its core idea is to maximize the mutual information (MI) between aligned entities' representations across different graphs while preserving attribute and structural information within each graph. This process uncovers complementary semantics across graphs to obtain more robust entity vectors. In practice, MIKG can be implemented using PyTorch/TensorFlow + DGL. Hyperparameters include entity vector dimension, number of selected attributes K, number of GAT layers, InfoNCE temperature $\tau$, and loss weight $\lambda$. Training employs Adam optimization, mini-batch negative sampling, and early stopping, enabling scalability to large-scale KGs.

The actual implementation process can be summarized as shown in Fig. 2:

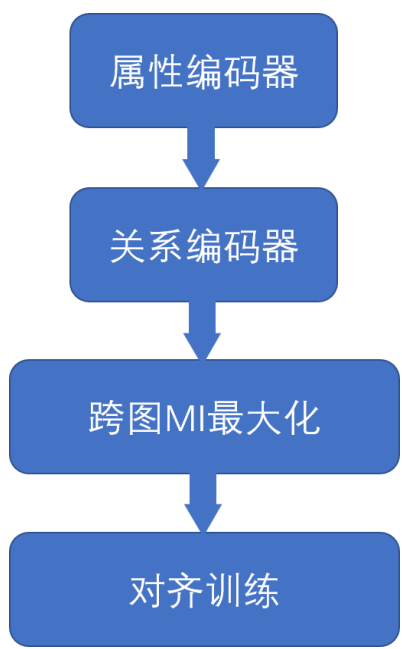

Fig.2. Implementation Process for Method B

## 4.1. Strong Points

**Strengths of Multi-KG4Rec Implementation (S1-S3)**

S1 Unifies the propagation of multimodal features within the KG structural space, eliminating the need for multi-model architectures;

S2 The graph attention mechanism adaptively assigns weights to neighbor relationships, balancing structural and content considerations;

S3 Demonstrates superior performance to single-structure or text/visual methods across multiple datasets (MovieLens, Amazon).

## 4.2. Weak Points

**Weaknesses of Multi-KG4Rec Implementation (W1-W3)**

W1 Fails to interact at the modal (text–image) level, relying solely on structural propagation;

W2 Insufficient integration of textual or visual features for cold-start items/isolated nodes;

W3 Deep neighborhood expansion in graph neural networks leads to excessive smoothing, necessitating more effective regularization or residual structures.

## 4.3. Detailed Analysis

(1) Graph Attention Networks only assign weights at the node level; it is recommended to introduce edge-level or triplet-level attention to capture fine-grained relationships between entities;

(2) Cold-start experiments only report overall performance without separately analyzing gains for cold-start nodes;

(3) Key hyperparameters such as the number of attention heads and layers have not been thoroughly investigated for their trade-offs between effectiveness and efficiency.

## 5. Proposed Method

By synthesizing the strengths and weaknesses of existing literature methods, we propose a "Cross-Graph Cross-Modal Mutual Information-Driven Unified Knowledge Graph Learning and Recommendation Framework" (CrossGMMI-DUKGLR). This approach integrates the multimodal fusion philosophy of Multi-KG4Rec with the cross-graph mutual information maximization strategy of MIKG. It aims to simultaneously address entity alignment and information sharing across knowledge graphs, joint modeling of multimodal data both within and across graphs, as well as noise robustness, efficient negative sampling, and large-scale scalability.

The innovation of this model lies in its ability to simultaneously maximize cross-map mutual information between attribute/text and structural representations, enhancing information sharing after alignment. Compared to MIKG, which aligns only structure or attributes, this multimodal collaboration yields richer semantics. Building upon Multi-KG4Rec, we incorporate a Cross-Attention mechanism to achieve deep interaction among text, image, and structure. Additionally, we employ a memory bank combined with random sampling techniques for contrastive learning negative sample generation, ensuring more stable and efficient training. Integrated with Jumping-Knowledge and graph transformation augmentation, this approach enhances robustness against noise and long-range dependencies, potentially supporting knowledge graphs at the million-entity scale. CrossGMMI-DUKGLR simultaneously addresses cross-graph entity alignment and intra-graph multimodal deep fusion. By introducing efficient dynamic negative sampling and graph transformation augmentation, it not only enhances entity alignment robustness but also further improves personalized recommendation quality. This approach demonstrates clear innovation and industrial implementation potential.

## 5.1. Overall Design

This method first performs unified preprocessing on knowledge graphs from diverse sources through preliminary entity alignment and redundancy noise reduction. It then constructs cross-graph subgraphs based on a sampling strategy, incorporating both multimodal information and structural subgraphs of n-hop neighbors. Building upon this foundation, the system employs parallel multimodal and structural encoders for feature extraction: attribute text is encoded by pre-trained BERT and aggregated via self-attention; image features are extracted using CLIP and deeply fused with text representations through cross-attention; structural information is dynamically weighted and aggregated for neighbors and relationships using enhanced graph neural networks like GAT+Jumping-Knowledge.

On top of the multimodal and structural vectors output by the encoder, we introduce a cross-graph mutual information contrastive learning module. By constructing positive-negative sample pairs for the same entity across different graphs, we maximize the mutual information estimated by the InfoNCE loss to achieve self-supervised entity alignment and knowledge sharing. After pre-training, the aligned multimodal and structural representations are concatenated into a unified vector. This vector is then fine-tuned using cross-entropy or ranking loss to handle user-item interactions. This approach enhances recommendation accuracy while ensuring dynamic negative sampling efficiency and online incremental updating capabilities across million-scale graphs. This design balances robustness in cross-graph multimodal alignment while meeting industrial scalability and flexibility through two-stage training and a modular architecture.

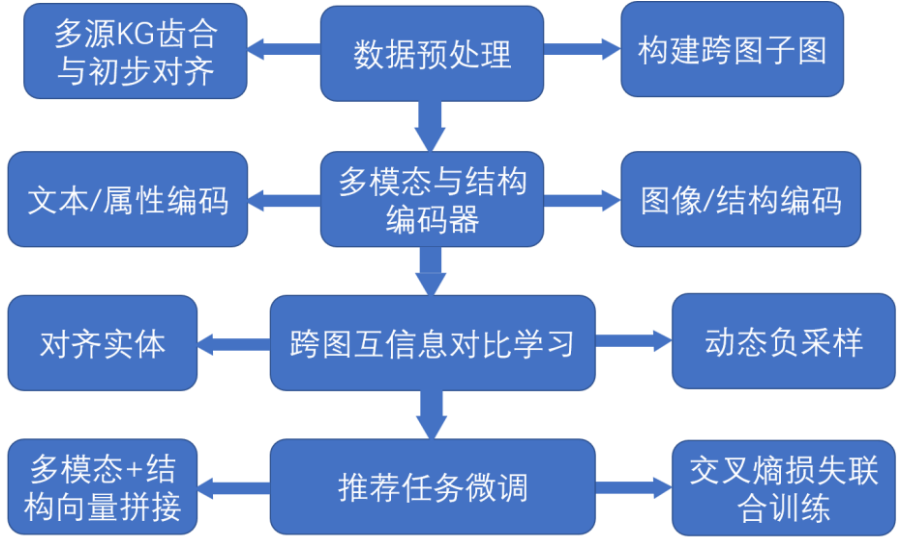

Fig. 3. CrossGMMI-DUKGLR Methodology Framework

## 5.2. Dataset Collection

This study utilizes two publicly available datasets: the cross-lingual entity alignment dataset DBP15K and the classic movie recommendation dataset MovieLens-1M. DBP15K directly extracts synonymous entity pairs from DBpedia, retaining entity names, English summaries, property triples, and interlink tags as positive examples for entity alignment, while unlabeled pairs serve as negative examples. MovieLens-1M contains rating records from 60,000 users for over 4,000 films. We classify ratings ≥ 4 as positive feedback, ratings ≤ 2 as negative feedback, and exclude data with ratings = 3. To incorporate multimodal and structural information, we utilized the TMDB/IMDB API to extract poster images, plot summaries, and relationship triples for movie entities in MovieLens. These were then mapped to KG nodes using unified IDs.

For feature selection, in the entity alignment task, we retain only the top K=10 most frequent relationship attributes (e.g., category, country, language) and use BERT to extract name and summary text vectors. For the recommendation task, we augment the structural features with CLIP poster features and plot text vectors. Attribute text is truncated to the first 256 words, and images are uniformly resized to 224×224 inputs. For the relationship subgraph, we sampled 2-hop neighbors to construct local structures. Regarding labels, DBP15K uses the provided alignment pairs as 1/0 classification labels; in the MovieLens task, positive and negative samples are generated by binarizing actual ratings. During training, we further supplement non-interaction pairs via random negative sampling to support final alignment and recommendation evaluation.

## 5.3. Model and Algorithm

The model architecture and algorithmic workflow can be divided into two phases: the "pre-training phase" and the "fine-tuning phase." The following outlines the primary modules and information flow:

Encoder Module

Text/Attribute Encoder：$h_i^T = BERT(attr\_i)$

Image Encoder：$h_i^I = CLIP(img\_i)$

Cross-Attention Integration：

$Q = W\_q\ h_i^T$，$K = W\_k\ h_i^I$，$V = W\_v\ h_i^I$
$H\_i = softmax(QK^T/\sqrt{d})\ V$

Structural Encoder (L-layer GNN + Dropout Connections)：

$\alpha\_{ij} = softmax\_j(e\_{ij})$，$e\_{ij}=a^T[W\_h\ h_i^l\ //\ W\_r\ r\_{ij}\ //\ W\_h\ h_j^l]$
$h_i^{l+1} = \sigma(\sum\{j \in N(i)\}\ \alpha\{ij\} \cdot h\_j^l) + Jump(h_i^l)$

Mutual Information Contrastive Learning

The two representations of the same entity i in KG_A and KG_B are denoted as $z_i^A$ and $z_i^B$, respectively. The positive-negative sample comparison loss is InfoNCE:

$L\_MI = -\sum\{i=1\}^N \log \exp(sim(z_i^A, z_i^B)/\tau)\ /\ \sum\{j=1\}^K \exp(sim(z_i^A, z_j^B)/\tau)$

Where $sim(u,v) = u^T\ v\ /\ \|u\|\ \|v\|$. By minimizing L_MI, the model brings multimodal + structural representations across entities in the graph closer together in the vector space.

Recommended Fine-Tuning

For user u and item v, obtain vectors $h\_u$ and $h\_v$ respectively from the above encoders, and predict the rating $s(u,v) = h\_u^T\ h\_v$. Train using binary cross-entropy or BPR loss:

L_rec = −[y·logσ(s)+(1−y)·log(1−σ(s))]

The pseudocode is as follows:

Input: KG_A, KG_B, interaction data D = {(u, v, y)}, hyperparameters: learning rate η, τ, number of GNN layers L

Initialization: parameters Θ

// Pre-training phase

for epoch in 1…E1:

for entity pair {(i^A,i^B)} in batch:

# Encoding

z_i^A = Encoder(i^A; Θ), z_i^B = Encoder(i^B; Θ)

# Compute InfoNCE loss

L_MI = −∑ log exp(sim(z_i^A,z_i^B)/τ) / ∑_neg exp(sim(z_i^A,z_neg)/τ)

Θ ← Θ − η∇_Θ L_MI

// Fine-tuning phase

for epoch in 1…E2:

for (u,v,y) in D batches:

h_u = Encoder(u; Θ), h_v = Encoder(v; Θ)

s = h_u^T h_v

L_rec = −[y·logσ(s)+(1−y)·log(1−σ(s))]

Θ ← Θ − η∇_Θ L_rec

Output: Model parameters Θ

CrossGMMI-DUKGLR consists of four main components that work synergistically to generate personalized recommendations:

(1) Cross-Modal Feature Extractor:

For each modality m$^{(k)}$, we employ a specialized encoder $f_k$ to extract features:

$$h_i^{(k)} = f_k(m_i^{(k)}; \theta_k)$$

where $\theta_k$ represents the parameters of encoder k. For textual modalities, we use BERT-based encoders; for visual modalities, we employ ResNet or Vision Transformer architectures; for audio, we utilize wav2vec 2.0.

(2) Knowledge Graph Construction:

We initialize entity embeddings $e_i \in R^d$ for all entities in the knowledge graph using the multimodal features:

$$e_i = g(h_i^{(1)}, h_i^{(2)}, ..., h_i^{(k)}; \emptyset)$$

Where $g$ is a learnable aggregation function parameterized by $\phi$. We experiment with various aggregation strategies including concatenation with dimensionality reduction, attention-based pooling, and gated fusion.

(3) Multimodal Fusion:

The multimodal fusion module adaptively combines features from different modalities based on user preferences and item characteristics. We introduce a personalized fusion mechanism that learns user-specific and item-specific fusion weights:

$$w_u^{(k)} = \text{softmax}(W_u p_u + b_u)$$

$$w_i^{(k)} = \text{softmax}(W_i q_i + b_i)$$

Where $p_u$ and $q_i$ are user and item profile embeddings learned from interaction history.

(4) Personalized Recommendation:

The final recommendation score is computed by combining the fused multimodal representation with knowledge graph reasoning:

$$y_{ui} = \sigma(v^T[z_{ui} \oplus path_{ui}])$$

Where $pathui$ represents features extracted from reasoning paths between user $u$ and item $i$ in the knowledge graph, and $\oplus$ denotes concatenation.

## 6. Discussions and Future Work

This method demonstrates distinct advantages in cross-knowledge-graph and multimodal deep integration: On one hand, it achieves self-supervised alignment through maximizing cross-graph mutual information, enabling the sharing of rich semantic information across different graphs without requiring manual annotation. On the other hand, the introduction of tri-modal encoding (text–image–structure) and dynamic negative sampling not only enhances the robustness of alignment and recommendation tasks but also ensures the model's scalability and online incremental update capability under million-scale knowledge graphs. Furthermore, the two-stage training strategy (pre-training alignment → downstream fine-tuning) improves alignment accuracy while effectively balancing recommendation performance. However, this approach also has potential limitations, including high computational overhead and strong dependencies on hyperparameters and pre-trained models during pre-training and fine-tuning. Its generalizability and real-time capability, particularly in resource-constrained environments or emerging knowledge graph domains, remain to be validated.

Future improvements and expansions can be pursued in the following areas: First, introducing more efficient mutual information estimation techniques and lightweight model distillation to reduce the computational cost of cross-modal encoding and contrastive learning. Second, exploring joint modeling of graph dynamic evolution and temporal information to enhance adaptability in time-sensitive scenarios. Third, integrating privacy-preserving and federated learning mechanisms into cross-graph alignment frameworks to address collaborative analysis of sensitive data. Finally, extending this approach to practical applications involving multi-knowledge-source fusion such as medical diagnosis and financial risk control to achieve a closed-loop process from knowledge graph alignment to decision support. Extending the framework to handle dynamic multimodal content, investigating more sophisticated personalization mechanisms that consider temporal factors, and developing efficient approximation techniques for large-scale deployment. We also plan to conduct user studies to evaluate the quality of generated explanations and their impact on user trust and satisfaction.

## 7. References

[1] Brett Karlan, Human achievement and artificial intelligence. *Ethics. Inf. Technol.* 2023, **25**: 40.

[2] Fan Ouyang, Mian Wu, Luyi Zheng, et al, Integration of artificial intelligence performance prediction and learning analytics to improve student learning in online engineering course. *Int. J. Educ. Technol. High. Educ.* 2023, **20**: 4.

[3] Erik Hermann, Stefano Puntoni, Artificial intelligence and consumer behavior: From predictive to generative AI. *J. Bus. Res.* 2024, **180**: 114720.

[4] Qingzong Li, Pingyu Jiang, Jianwei Wang, et al, A kind of intelligent dynamic industrial event knowledge graph and its application in process stability evaluation. *J. Intell. Manuf.* 2025, **36**: 1801-1818.

[5] Jože M. Rožanec, Blaž Fortuna, Dunja Mladenić, Knowledge graph-based rich and confidentiality preserving explainable artificial intelligence (XAI). *Inform. Fusion* 2022, **81**: 91-102.

[6] Zenglong Wang, Xuan Liu, Zheng Liu, et al, A link prediction method for multi-modal knowledge graphs based on adaptive fusion and modality information enhancement. *Neural Networks* 2025, **191**: 107771.

[7] Kang Yang, Ruiyun Yu, Bingyang Guo, et al, Is multi-level data enhancement helpful for knowledge graph? A new perspective on multimodal fusion. *Knowl-Based Syst.* 2024, **301**: 112285.

[8] Gurinder Kaur, Fei Liu, Yi-Ping Phoebe Chen, Engineering applications of artificial intelligence. *Eng. Appl. Artif. Intel.* 2024, **135**: 108792.

[9] L. He, W. Ye, Y.X. Wang, et al, using knowledge graph and *RippleNet* algorithms to fulfill smart recommendation of water use policies during shale resources development. *J. Hydrol*. 2023, **617**: 128970.

[10] Wei Zhang, Ling Kong, Soobin Lee, et al, detecting mental and physical disorders using multi-task learning equipped with knowledge graph attention network. *Artif. Intell. Med.* 2024, **149**: 102812.

[11] Kian P. Abdolazizi, Roland C. Aydin, Christian J. Cyron, et al, Constitutive kolmogorov– arnold networks (CKANs): Combining accuracy and interpretability in data-driven material modeling. *J. Mech. Phys. Solids* 2025, **203**: 106212.

[12] Xiang Wang, Tinglin Huang, Dingxian Wang, et al, Learning intents behind interactions with knowledge graph for recommendation. *arXiv e-prints* 2021, **21**: 19-23.

[13] Ping Yang, Wengxiang Chen, Hang Qiu, MMGCN: Multi-modal multi-view graph convolutional networks for cancer prognosis prediction. *Comput. Meth. Prog. Bio*. 2024, **257**: 108400.

[14] Peizhang Li, Qing Fei, Zhen Chen, Interpretable multi-agent reinforcement learning via multi-head variational autoencoders. *Appl. Intell.* 2025, **55**: 577.

[15] Zhitao Wang, Wenjie Li, Hanjing Su, Hierarchical Attention Link Prediction Neural Network. *Knowl-Based Syst.* 2021, **232**: 107431.

[16] Zihan Wei, Ning Wu, Fengxia Li, et al, MoCo4SRec: A momentum contrastive learning framework for sequential recommendation. *Expert. Syst. Appl.* 2023, **223**: 119911.

[17] Jingjing Wang, Haoran Xie, Siyu Zhang, et al, Multimodal fusion framework based on knowledge graph for personalized recommendation. *Expert. Syst. Appl*. 2025, **268**: 126308.

[18] Ao Gaoa, Mingda Li, Wei Liu, et al, Maximizing mutual information across knowledge graphs for robust entity alignment. *Expert. Syst. Appl*. 2026, **296**: 129031.